\documentclass[pra,superscriptaddress,twocolumn,showpacs,preprintnumbers,amsmath,amssymb,showkeys]{revtex4}
\usepackage{graphicx}
\usepackage{dcolumn}
\usepackage{bm}
\begin{document}
\title{Bose-Einstein condensation of photons in an ideal atomic gas}

\author{Alex Kruchkov}
\affiliation{Karazin Kharkiv  National University, 4 Svobody Sq., 61077 Kharkiv, Ukraine}
\email{aleks.kryuchkov@gmail.com}

\author{Yurii Slyusarenko}
\affiliation{Karazin Kharkiv National University, 4 Svobody Sq., 61077 Kharkiv, Ukraine}
\affiliation{Akhiezer Institute for Theoretical Physics, NSC KIPT, 1 Akademichna Str., 61108 Kharkiv, Ukraine}
\email{slusarenko@kipt.kharkov.ua}

 \noindent
 \noindent
\date{\today}
\begin{abstract}
We study peculiarities of Bose-Einstein condensation of photons that are in thermodynamic equilibrium with atoms of non-interacting gases.General equations of the thermodynamic equilibrium of the system under study are obtained. We examine solutions of these equations in the case of high temperatures, when the atomic components of the system can be considered as nondegenerated ideal gases of atoms, and photonic component can form a state with Bose condensate. Transcendental equation for transition temperature and expression for the density of condensed photons in considered system  are derived. We also obtain analytical solutions of the equation for the critical temperature in a number of particular cases. The existence of two regimes of Bose condensation of photons, which differ significantly in nature of transition temperature dependence on the total density of  pumped into the system photons, is revealed. In one case, this dependence is a traditional fractional-power law, and in another one it is the logarithmic law. Applying numerical methods, we determine boundaries of existence and implementation conditions for different regimes of condensation depending on the physical parameters of the system under study. We also show that for a large range of physical systems that are in equilibrium with photons (from ultracold gases of alkali metals to certain types of ideal plasma), the condensation of photons should occur according to the logarithmic regime.
\end{abstract}



\maketitle

%
\section{Introduction}
\label{intro}

The phenomenon of Bose-Einstein condensation (BEC) became nowadays so accustomed that the possibilities of its wide applications are seriously discussed. It became reality due to implementation of BEC in various many-particle systems, from ultracold gases of atoms and molecules (see e.g. \cite{1}-\cite{5}) to exciton-polaritons (e.g. \cite{polariton1}, \cite{polariton2}). As is well known, the basis of BEC phenomenon and related phenomenon of superfluidity is the property of a macroscopic number of bosons to occupy the same quantum state. It is also well known that such a property of bosons leads to a specific form of the distribution function. It was first noted by A. Einstein\cite{6} after reading the work of S. Bose \cite{7}. Recall that in mentioned paper S. Bose was the first to derive an expression for the photon distribution function. That is, the energy distribution function of bosons was originally derived for photons, but not for Bose atoms. For this reason, the idea of observation of photon BEC seems quite obvious. However, until recently one considered to be impossible to carry out experimentally the conditions for such condensation to be achieved. The reasons for such a statement seem to have been caused by several fairly obvious circumstances. In the case of a system consisting of Bose atoms, the possibility of BEC phase appearance with temperature decreasing can be shown even in the ideal gas model. In this case, Bose condensate is formed by atoms with zero momentum. However it is known that in vacuum photons with zero momentum do not exist. It reflects currently accepted fact that photon is considered to be massless in vacuum. Besides, it is difficult to imagine the method to lower the temperature in a weakly nonideal gas of photons. If one considers photons in a medium with decreasing temperature, the number of photons is reduced due to absorption by substance. But one of the BEC transition conditions is the conservation of the total number of particles, that is, the non-condensate particles and those ones forming  Bose condensate.

	It should be noted that there were a few works of the last century (see  e.g. \cite{8}-\cite{10}), which in some way touched upon the issues related to Bose condensation of photons. These works had risen despite of so obvious (as it would seem) arguments in favour of photons condensation impossibility. In \cite{8} authors hypothesized the existence of a photon Bose condensate in early stages of the Universe expansion, assuming a non-zero photon mass. Authors of \cite{8} also considered the possibility that condensate evaporation due to interaction with charged particles leads to the formation of longitudinal relict radiation. It was assumed that this radiation forms the main part of Universe mass. In \cite{9} authors studied the process of establishing equilibrium between radiation and matter in fully ionized plasma. Solution of kinetic equation indicated the possibility of photon BEC formation in the system without photon absorption. Both the scattering processes (for example, photons by electrons) and the processes of emission and absorption were considered to be responsible for the relaxation mechanism. However, it was argued in \cite{9} that the presence of photon absorption leads to the replacement of photons condensation phenomenon to its temporary accumulation at the region of low frequencies. In \cite{10} authors studied both the possibility of formation of effectively two-dimensional photonic superfluid in Fabry-Perot cavity, and the existence of Bogoliubov elementary excitations in the system under study, as well as the methods of experimental observation of such superfluidity, based on Landau criterion.

Until recently such works were considered to be rather "exotic". The situation had dramatically changed after appearance of the article \cite{11} (see also \cite{12}), in which  the implementation of BEC of photons in a real experiment at room temperature was announced. In this work, the photons are not allowed to leave the system (that is a special dye) with the help of specifically placed mirrors with extremely high reflectivity. The system was pumped by photons with the help of external laser. Relaxation mechanism is determined by multiple absorption of laser photons by dye molecules and re-emission of a lower energy photons. Thus it became possible to achieve the equilibrium state of photons and environment at room temperature without photons loss. Further pumping photons into the system led to the increase of the density of  "free" photons that created the possibility of its condensation. Note that in mentioned paper the system under study is also considered to be effectively two-dimensional. As for the mass of the photon, the calculations of authors \cite{11} gives numerical value approximately equal $6.7\cdot10^{-33}  g$.

It is effective mass of photons we are talking about. It appears due to a significant change of the photon dispersion law in matter as compared with its linear dispersion law in vacuum. As is well known, the spectrum of photons in matter reveals so-called 'cut-off frequency', that is the final value of the frequency at zero wave vector. The existence of a finite cut-off frequency is analogous to the presence of an effective mass in the photon dispersion law (see further). We emphasize that the appearance of cut-off frequency in spectrum and, therefore, the effective mass of photon, is caused by its interaction with matter. In this sense photon in matter can be considered as a quasiparticle. Note also that photon effective mass can appear as a result of boundary conditions. For example, if the system is situated between two concave mirrors with high reflectivity, as it is in case of real experiment \cite{11}-\cite{12}. The presence of these mirrors does not only conserve the total number of photons in system and make it effectively two-dimensional, but also contributes to the establishment of standing wave along a perpendicular to the mirrors' surface. The latter circumstance leads to the appearance of frequency cut-off in two-dimensional photonic spectrum (see in this regard \cite{10}).

Thus, the conditions of the experiment \cite{11} have eliminated all mentioned above difficulties , that were earlier preventing to realize the photon BEC experimentally and to implement the "scientific sensation".

However, an achieved breakthrough enables, in our opinion, to go beyond the framework of photon BEC studies in effectively two-dimensional systems, and to look for the models of other physical systems, where the implementation of photon Bose condensation is also available in the region of relatively high temperatures. In the present paper, in this regard, we have determined conditions for Bose-Einstein condensation of photons which are in thermodynamic equilibrium with the atoms of diluted gases. Special attention is paid for the case of high temperatures, when rarefied atomic gases can be regarded as nondegenerate.

\section{General equations of thermodynamic equilibrium of photons and atoms of diluted gas}
\label{general equations}

To solve this problem it is proposed to proceed from the following model. Consider the system to be composed of an ideal gas of Fermi or Bose atoms that are in thermodynamical equilibrium with photons. Atoms of this gas can be only in two quantum states: the ground one and the excited one (two-level atoms). In other words, the transition between one atomic state to another one happens as a result of absorption or emission of a photon. Thus, the excited atom can be regarded as a bound state of a photon and an atom in the ground state (in this connection see \cite{13}). The possibility of considering the system as an ideal gas of atoms and photons is based on the assumption that once thermodynamic equilibrium is reached in a system, the interaction between its components can be neglected.

From a formal (mathematical) point of view, we can consider the atomic components of the system to be composed by two kinds of atoms that are in equilibrium with photon gas.
We assume that type "1" of atoms is characterized by a set of quantum numbers $\alpha _{1} $ and the distribution function of atoms

\begin{equation}\begin{split}\begin{gathered}
 \label{1}
f_{\alpha _{1} } \left(\textbf{\textit{p}}\right)=\left\{\exp \left[\frac{\varepsilon _{\alpha _{1} } \left(\textbf{\textit{p}}\right)-\mu _{1} }{T} \right]\pm 1\right\}^{-1},
\\
\varepsilon _{\alpha _{1} } \left(\textbf{\textit{p}}\right)\equiv \varepsilon _{\alpha _{1} } +\frac{\textbf{\textit{p}}^{2} }{2m},
\end{gathered}\end{split}\end{equation}

\noindent ($\varepsilon _{\alpha _{1} } $is the energy of levels of an atom at rest with a set of quantum numbers $\alpha _{1} $, $\varepsilon _{\alpha _{1} } <0$, $\mu _{1} $ is the chemical potential, $m$ is the mass of atom) and the type "2" of atoms is characterized by a set of quantum numbers $\alpha _{2} $ and the distribution function of atoms

\begin{equation}\begin{split}\begin{gathered}
 \label{2}
f_{\alpha _{2} } \left(\textbf{\textit{p}}\right)=\left\{\exp \left[\frac{\varepsilon _{\alpha _{2} } \left(\textbf{\textit{p}}\right)-\mu _{2} }{T} \right]\pm 1\right\}^{-1} ,
\\
\varepsilon _{\alpha _{2} } \left(\textbf{\textit{p}}\right)\equiv \varepsilon _{\alpha _{2} } +\frac{\textbf{\textit{p}}^{2} }{2m},   \end{gathered}\end{split}\end{equation}

\noindent where $\varepsilon _{\alpha _{2} } $ is the energy of levels of an atom at rest with a set of quantum numbers $\alpha _{2} $, $\varepsilon _{\alpha _{2} } <0$, and $\mu _{2} $ is the chemical potential. In formulae \eqref{1} - \eqref{2}, the sign "$-$" must be chosen in the case of bosonic gas of atoms, and the sign "$+$" in the case of fermionic gas. For definiteness, we assume that the set of quantum numbers   $\alpha _{1}$  corresponds to the ground state of the atom, and a set of quantum numbers  $\alpha _{2}$  corresponds to the excited state of the atom.

We consider also photons that are in thermodynamic equilibrium with two-component diluted gas to be described by the distribution function

\begin{equation}\begin{split}\begin{gathered}
 \label{3}
f_{ph} \left(\textbf{\textit{k}}\right)=\left\{\exp \left[\frac{\hbar \omega \left(\textbf{\textit{k}}\right)-\mu ^{*} }{T} \right]-1\right\}^{-1},
 \end{gathered}\end{split}\end{equation}

\noindent where $\omega \left(\textbf{\textit{k}}\right)$ is the dispersion law of photons and $\mu ^{*} $ is their chemical potential. The introduction of non-zero chemical potential of photons implies the conservation of their total number in a given thermodynamic state of system. In the formulae (1) - (3) and throughout the text of the present paper one assumes that the temperature of the system $T$ is measured in energy units.

Condition of thermodynamic equilibrium means that at given temperature in system, the atomic $N$ and the photonic $N_{ph} $ total numbers are conserved:

\begin{equation}
 \label{4}
N=g_{\alpha _{1} } \sum _{\textbf{\textit{p}}}f_{\alpha _{1} } \left(\textbf{\textit{p}}\right) +g_{\alpha _{2} } \sum _{\textbf{\textit{p}}}f_{\alpha _{2} } \left(\textbf{\textit{p}}\right),
\end{equation}

\begin{equation}
 \label{5}
N_{ph} =g^{*} \sum _{\textbf{\textit{k}}}f_{ph} \left(\textbf{\textit{k}}\right) +g_{\alpha _{2} } \sum _{\textbf{\textit{p}}}f_{\alpha _{2} } \left(\textbf{\textit{p}}\right),
\end{equation}

\noindent where $g_{\alpha _{1} } $, $g_{\alpha _{2} } $ are parameters that take into account the degeneracy multiplicity of atomic levels with the set of quantum numbers $\alpha _{1} $, $\alpha _{2} $ respectively. For example, if the energy level with a set of quantum numbers $\alpha _{1} $ is spin-degenerated, then $g_{\alpha _{1} } =2S_{\alpha _{1} } +1$. Similarly, the quantity $g^{*} $ takes into account the possible energy degeneracy of a photon with energy $\hbar \omega \left(\textbf{\textit{k}}\right)$. For example, if this degeneracy is associated only with photons polarization, the sum over polarizations gives $g^{*} =3$. Recall that in the presence of matter photons can have a longitudinal polarization, and it was taken into account in the previous statement. We emphasize that equation \eqref{5} reflects the conservation of the total number of pumped into system photons, that is, the total amount of free photons and photons absorbed by two-level atoms.
It is necessary to make here the following remark. In principle regard, the existence of mixed states of atoms, resulting from their interaction with photons, is possible. However, the description of the system in means of distribution functions \eqref{1} -\eqref{3} refers to the use of the Wigner (in fact, quasi-classical) approximation in statistical mechanics. This level of description cannot allow taking in consideration the influence of mixed states of atoms on the behavior of the system. Therefore in the present study the possibility of the existence of such states is neglected, which is reflected in equations \eqref{4} and \eqref{5}. Clarification on the issue of contribution of the mixed states of atoms in thermodynamics of system requires further research.

 Since atomic states $\alpha _{1} $, $\alpha _{2} $ differs in single-photon transition, then chemical potentials $\mu _{1} $, $\mu _{2} $ and $\mu ^{*} $ are supposed to be connected by following equation (so-called Gibbs rule or chemical reaction condition, see in this regard \cite{13} and \cite{14}):

\begin{equation}
 \label{6}
\mu _{1} +\mu ^{*} =\mu _{2}.
\end{equation}

Equations \eqref{4} - \eqref{6} form a complete system of coupled equations to describe thermodynamic equilibrium between photons and ideal gas of two-level bosons or fermions. Deriving from these equations chemical potentials as  functions of temperature, total number of photons and total number of atoms, thus we determine in accordance with  \eqref{1} - \eqref{3} the distribution functions of all the three components of the system under study.

 However, as is easily seen, before solving the problem it is necessary to clarify the specific type of photons dispersion law $\omega \left(\textbf{\textit{k}}\right)$. As we mentioned above, the type of dispersion law determines the effective mass of photons in matter, thus determining the conditions of their Bose condensation. In the present paper, following \cite{9}-\cite{12}, we assume photon dispersion law to be quadratic on the wave vector (or photon momentum). It can be 'justified' if one starts from the expression for energy of the photon in matter as a relativistic object \cite{10}-\cite{12}:

\begin{equation}
 \label{7}
\hbar \omega \left(\textbf{\textit{k}}\right)=\hbar \sqrt{\omega _{0}^{2} +v^{2} k^{2} },
\end{equation}

\noindent where $v$ is the speed of light in matter and $\omega _{0} $ is the spectrum cut-off frequency. Then in the region of small wave vectors one obtains

\begin{equation}\begin{split}\begin{gathered}
 \label{8}
\hbar \omega \left(\textbf{\textit{k}}\right)=\hbar \sqrt{\omega _{0}^{2} +v^{2} k^{2} } \approx \hbar \omega _{0} \left(1+\frac{1}{2} \frac{v^{2} k^{2} }{\omega _{0}^{2} } \right)
\\
=\hbar \omega _{0} +\frac{v^{2} \left(\hbar k\right)^{2} }{2\hbar \omega _{0} } =\hbar \omega _{0} +\frac{\left(\hbar k\right)^{2} }{2m^{*} },
\end{gathered}\end{split}\end{equation}

\noindent which implies that photons with momentum $\hbar \textbf{\textit{k}}$ possess an effective mass $m^{*} $

\begin{equation}
 \label{9}
m^{*} \equiv \frac{\hbar \omega _{0} }{v^{2} }.
\end{equation}

\noindent As was already mentioned, the presence of cut-off frequency $\omega _{0} $ in photon spectrum may be caused by different reasons, both related to the interaction of photons with matter, and the presence of external constraints such as mirrors. In this sense, the shape of spectrum \eqref{7} is not dictated by "relativism" in any way. In fact, as is well known, the dispersion equations for electromagnetic waves in plasma yield the following dispersion law for transverse waves (see e.g. \cite{15}):

\begin{equation}
\label{10}
\omega _{t}^{2} \left(\textbf{\textit{k}}\right)=\omega _{0}^{2} +c^{2} k^{2},
\end{equation}

\noindent where $c$ is the speed of light in vacuum, and $\omega _{0} $ is the plasma frequency (the frequency of Langmuir oscillations)

\begin{equation}
\label{11}
\omega _{0}^{2} =\frac{4\pi ne^{2} }{m}.
\end{equation}

\noindent In the last formula, $n$ is the density of electrons in plasma and $e$ is the elementary charge. Note that longitudinal waves in plasma also have a quadratic dispersion law

\begin{equation}
\label{12}
\omega _{l} \left(\textbf{\textit{k}}\right)=\omega _{0} \left(1+\frac{3}{2} \left(kr_{D} \right)^{2} \right),
\end{equation}

\noindent where $r_{D} $ is the radius of Coulomb screening (Debye radius)

\begin{equation}
\label{13}
r_{D} =\left(\frac{T}{8\pi ne^{2} } \right)^{1/2},
\end{equation}

\noindent One can see that in the case of fully ionized plasma the existence of cut-off frequency is not caused by object's relativistic nature at all.

 Naturally, the consistent description requires the calculation of cut-off frequency for the photons' dispersion law and determining the speed of propagation of electromagnetic waves in the system under study. However, both the derivation of dispersion equations for specific mediums, and their solutions are separated problems, and solving each of them involves considerable mathematical difficulties. Such difficulties, that rise while deriving and solving the dispersion equations for electromagnetic waves in the systems with bound states and in the presence of a Bose condensate, can be estimated from works \cite{16}-\cite{19}. Therefore, in this paper we shall not intend to determine the cut-off frequency of the photon spectrum and their velocities in the present system, considering them to be given. Moreover, in all subsequent calculations, the energy of the photons (see \eqref{7}) to be written as:

\begin{equation}
\label{14}
\hbar \omega \left(\textbf{\textit{k}}\right)\equiv \hbar \omega \left(\textbf{\textit{p}}\right)=\hbar \omega _{0} +\frac{\textbf{\textit{p}}^{2} }{2m^{*} },
\end{equation}

\noindent where the notation for the photon momentum $\textbf{\textit{p}}=\hbar \textbf{\textit{k}}$ was introduced and the quantity $m^{*} $ is given by \eqref{9}.

It is necessary to give here some clarification. Since 2006 \cite{polariton1}, studies related to the phenomenon of BEC in an exciton-polariton systems are actively conducting. Sufficiently detailed information on the state of affairs in this research area can be found in \cite{polariton2}, also taking into account available links therein. In this regard, we emphasize that in the present paper we are talking about the photons in matter instead of polaritons of any kind. One can easily convince in it after recalling that polaritons are quasi-particles, combining photons and elementary excitations in the system, caused by the presence of interaction between the structural units of matter. For example, the exciton-polaritons are bound states of photons and excitons. It is known that the introduction of the concept of elementary excitations quanta  makes sense if there is a strong interaction between the structural units of matter. For this reason in \cite{11}-\cite{12} authors specified that the conditions of a real experiment eliminate the existence of polaritons, and therefore it was the BEC of photons to be observed. In the present study we show the possibility of Bose condensation of photons in an ideal gas, i.e. in the system where interactions between particles can be neglected. Thus, as was already mentioned (see \eqref{7} - \eqref{14}), the photon spectrum in such substances may be suitable for formation in the photonic system the states with BEC, under certain conditions .

Hence, taking into account \eqref{14} and proceeding in \eqref{4} and \eqref{5} from the momentum sums to integrals according to the standard rules of quantum mechanics, we obtain the following general equations of thermodynamic balance of radiation and matter (ideal gas of atoms):

\begin{equation}\begin{split}\begin{gathered}
\label{15}
n=\frac{g_{\alpha _{1} } }{2\pi ^{2} \hbar ^{3} } \int _{0}^{\infty }dp \frac{p^{2} }{\exp \left[\frac{\varepsilon _{\alpha _{1} } -\mu _{1}+\left({p^{2} \mathord{\left/ {\vphantom {p^{2}  2m}} \right. \kern-\nulldelimiterspace} 2m} \right)}{T} \right]\pm 1}
\\
+\frac{g_{\alpha _{2} } }{2\pi ^{2} \hbar ^{3} } \int _{0}^{\infty }dp \frac{p^{2} }{\exp \left[\frac{\varepsilon _{\alpha _{2} } -\mu _{2} +\left({p^{2} \mathord{\left/ {\vphantom {p^{2}  2m}} \right. \kern-\nulldelimiterspace} 2m} \right)}{T} \right]\pm 1},
\end{gathered}\end{split} \end{equation}

\begin{equation}\begin{split}\begin{gathered}
\label{16}
n_{ph} =\frac{g^{*} }{2\pi ^{2} \hbar ^{3} } \int _{0}^{\infty }dp\frac{p^{2} }{\exp \left[\frac{\hbar \omega _{0} -\mu^{*}+\left({p^{2} \mathord{\left/ {\vphantom {p^{2}  2m^{*}}} \right. \kern-\nulldelimiterspace} 2m^{*}} \right)}{T} \right]-1}
 \\
+\frac{g_{\alpha _{2} } }{2\pi ^{2} \hbar ^{3} } \int _{0}^{\infty }dp \frac{p^{2} }{\exp \left[\frac{\varepsilon _{\alpha _{2} } -\mu _{2} +\left({p^{2} \mathord{\left/ {\vphantom {p^{2}  2m}} \right. \kern-\nulldelimiterspace} 2m} \right)}{T} \right]\pm 1},
\end{gathered}\end{split} \end{equation}

\begin{equation}
\label{17}
\mu _{1} +\mu^{*}=\mu _{2},
\end{equation}

\noindent  where $n$ and $n_{ph} $ are total densities of atoms or photons, respectively:

\begin{equation}\begin{split}\begin{gathered}
\label{18}
 n=\frac{N}{V} ,
\ \ \ \ \ \
 n_{ph} =\frac{N_{ph} }{V}
\end{gathered}\end{split} \end{equation}

\noindent  ($V$ is the volume of the system).

 It should be noted that in the framework of our model the values of $n_{ph}$ and $n$   should be in the same order of magnitude. This follows from the fact that we have declared a specific system of two-level atoms. This assumption is closer to reality if one considers

\begin{equation}
\label{19}
N_{ph} \sim N,
\ \ \ \ \ \
n_{ph} \sim n.
\end{equation}

\noindent   In this case one can hope that the number of excited atoms with the energy levels higher than $\varepsilon _{\alpha _{2} } $, is negligible.

Equations \eqref{15} - \eqref{18} should be considered as source equationes for solving assigned in the present article problem:  to determine the conditions under which there is a Bose condensate in a gas of photons that are in thermodynamic equilibrium with the atoms of dilute gases. Analytical solution of these equations is not generally possible. However, the most interesting case is the appearance of a Bose condensate of photons in the system under study in conditions when the atomic components are far from degeneration. In this case the system of equations \eqref{15} - \eqref{17} simplifies greatly, and it is possible to obtain some particular solutions in analytical form.

\section{Solution of thermodynamic balance equations for the system in symmetric phase}
\label{symmetric phase}

First of all, we study the solutions of equations \eqref{15} - \eqref{17} in the symmetric phase of the system far away from the transition point, when all the three components of the system can be considered as remote from degeneration. It will help us to determine relations between the system parameters that contribute to achieving the maximum possible number of free photons with temperature decreasing. Such behavior of the photon component would maximize the transition temperature of photons to a state with BEC. Recall that in the theory of phase transitions it is accepted to call the phase of higher symmetry just as "symmetric phase", and the phase of lower symmetry as "asymmetric phase".

To begin, we consider the case where all the three components are far from degeneration. This case is interesting due to the following circumstance: we are looking for the situation when in nondegenerated state the density of free photons increases with decreasing the temperature. This behavior of the photon component would effectively help in experimental realization of photonic BEC in the system under consideration. Conditions of nondegeneracy of all the three components can be written as:

\begin{equation}\begin{split}\begin{gathered}
\label{20}
\exp \left(\frac{\varepsilon _{\alpha _{1} } -\mu _{1} }{T} \right)\gg 1,
\ \ \ \ \
\exp \left(\frac{\varepsilon _{\alpha _{2} } -\mu _{2} }{T} \right)\gg 1,
\\
\exp \left(\frac{\hbar \omega _{0} -\mu^{*}}{T} \right)\gg 1.
\end{gathered}\end{split} \end{equation}

\noindent   Then momentum integrals from \eqref{15}, \eqref{16} can be calculated

\begin{equation}\begin{split}\begin{gathered}
\label{21}
\frac{g_{\alpha _{1} } }{2\pi ^{2} \hbar ^{3} } \int dp\frac{p^{2} }{\exp \left[\frac{\left({p^{2} \mathord{\left/ {\vphantom {p^{2}  2m_{1} }} \right. \kern-\nulldelimiterspace} 2m_{1} } \right)+\varepsilon _{\alpha _{1} } -\mu _{1} }{T} \right]\pm 1}
\\
\approx \exp \left[\frac {\mu _{1} -\varepsilon _{\alpha _{1} }} {T} \right]\left(\frac{mT}{2\pi \hbar ^{2} } \right)^{3/2} g_{\alpha _{1} },
\\
\frac{g_{\alpha _{2} } }{2\pi ^{2} \hbar ^{3} } \int dp\frac{p^{2} }{\exp \left[\frac{\left({p^{2} \mathord{\left/ {\vphantom {p^{2}  2M}} \right. \kern-\nulldelimiterspace} 2M} \right)+\varepsilon _{\alpha _{2} } -\mu _{2} }{T} \right]\pm 1}
\\
\approx \exp \left[\frac{\mu _{2} -\varepsilon _{\alpha _{2} }}{T} \right]\left(\frac{mT}{2\pi \hbar ^{2} } \right)^{3/2} g_{\alpha _{2} } ,
\\
\frac{g^{*}}{2\pi ^{2} \hbar ^{3} } \int _{0}^{\infty }dp\frac{p^{2} }{\exp \left[\frac{\hbar \omega _{0} -\mu^{*}+\left({p^{2} \mathord{\left/ {\vphantom {p^{2}  2m^{*}}} \right. \kern-\nulldelimiterspace} 2m^{*}} \right)}{T} \right]-1}
\\
\approx \exp \left[\frac{\mu^{*}-\hbar \omega _{0}}{T} \right]\left(\frac{m^{*} T}{2\pi \hbar ^{2} } \right)^{3/2} g^{*},
\end{gathered}\end{split} \end{equation}

\noindent and equations of  thermodynamic equilibrium \eqref{15} - \eqref{17} can be written as:

\begin{equation}\begin{split}\begin{gathered}
\label{22}
n=e^{\mu _{1} /T} g_{\alpha _{1} } \left(T\right)+e^{\mu _{2} /T} g_{\alpha _{2} } \left(T\right),
\\
n_{ph} =e^{\mu ^{*} /T} g^{*} \left(T\right)+e^{\mu _{2} /T} g_{\alpha _{2} } \left(T\right),
\\
e^{\mu _{1} /T} e^{\mu^{*}/T} =e^{\mu _{2} /T},
\end{gathered}\end{split}\end{equation}

\noindent where the following notations were introduced

\begin{equation}\begin{split}\begin{gathered}
\label{23}
g_{\alpha _{2} } \left(\frac{mT}{2\pi \hbar ^{2} } \right)^{3/2} \exp \left(-\varepsilon _{\alpha _{2} } /T\right)\equiv g_{\alpha _{2} } \left(T\right),
\\
g_{\alpha _{1} } \left(\frac{mT}{2\pi \hbar ^{2} } \right)^{3/2} \exp \left(-\varepsilon _{\alpha _{1} } /T\right)\equiv g_{\alpha _{1} } \left(T\right),
\\
g^{*} \left(\frac{mT}{2\pi \hbar ^{2} } \right)^{3/2} \exp \left(-\hbar \omega _{0} /T\right)\equiv g^{*} \left(T\right).
\end{gathered}\end{split}\end{equation}

\noindent Note that quantities $n_{\alpha _{1} } \left(T\right)$, $n_{\alpha _{2} } \left(T\right)$,

\begin{equation}\begin{split}\begin{gathered}
\label{24}
n_{\alpha _{1} } \left(T\right)=e^{\mu _{1} /T} g_{\alpha _{1} } \left(T\right),
\\
n_{\alpha _{2} } \left(T\right)=e^{\mu _{2} /T} g_{\alpha _{2} } \left(T\right),
\end{gathered}\end{split}\end{equation}

\noindent represent the density of atoms in quantum states $\alpha _{1} $ and $\alpha _{2} $, respectively, whereas the quantity  $\widetilde{n}_{ph} \left(T\right)$ is the density of free photons in the system:

\begin{equation}
\label{25}
\widetilde{n}_{ph} \left(T\right)=e^{\mu ^{*} /T}  g^{*}\left(T\right).
\end{equation}

\noindent where we have introduced the tilde sign  to avoid a confusion with the density of total pumped photons  $n_{ph}$.

Note also that in the case of nondegenerated atomic components, their statistics has a weak influence on the thermodynamic state of the system, since in the leading approximation both bosonic and fermionic distribution functions have the same (Boltzmann) form.

Solution of the system of equations \eqref{22} can be given as:

\begin{equation}\begin{split}\begin{gathered}
\label{26}
n_{\alpha _{1} } \left(T\right)=\frac{n}{\left[1+A\left(T\right)\right]},
\\
n_{\alpha _{2} } \left(T\right)=\frac{A\left(T\right)n}{\left[1+A\left(T\right)\right]},
\\
\widetilde{n}_{ph} \left(T\right)=\kappa \left(T\right) A\left(T\right),
\end{gathered}\end{split}\end{equation}

\noindent where, to simplify the form of expressions, we introduced following notations:

\begin{equation}\begin{split}\begin{gathered}
\label{27}
\kappa \left(T\right)=\frac{g^{*}\left(T\right)g_{\alpha _{1} } \left(T\right)}{g_{\alpha _{2} } \left(T\right)}
\\
=\frac{g^{*}g_{\alpha _{1} } }{g_{\alpha _{2} } } \exp \left[-\frac{\varepsilon _{\alpha _{1} } -\varepsilon _{\alpha _{2} } +\hbar \omega _{0} }{T} \right] \left(\frac{m^{*}T}{2\pi \hbar ^{2} } \right)^{3/2},
\end{gathered}\end{split}\end{equation}

\begin{equation}\begin{split}\begin{gathered}
\label{28}
A\left(T\right)\equiv \frac{1}{2} \left[\frac{n_{ph} -n}{\kappa \left(T\right)} -1\right]
\\
+\frac{1}{2} \left[1+\frac{\left(n_{ph} -n\right)^{2}}{\kappa^{2} \left(T\right)}  +2\frac{n_{ph} +n}{\kappa \left(T\right)} \right]^{1/2} .
\end{gathered}\end{split}\end{equation}

\noindent Consider the expression for the density of free photons (see (26)), taking into account expression \eqref{28}:

\begin{equation}\begin{split}\begin{gathered}
\label{29}
\widetilde{n}_{ph} \left(T\right)=\frac{1}{2} \left[\left(n_{ph} -n\right)-\kappa \left(T\right)\right]
\\
+\frac{1}{2} \left[\left(n_{ph} -n\right)^{2} +2\left(n_{ph} +n\right)\kappa \left(T\right)+\kappa ^{2} \left(T\right)\right]^{1/2} .
\end{gathered}\end{split}\end{equation}

\noindent It is clear that the temperature dependence $n_{ph} \left(T\right)$ is determined by a single parameter $\kappa \left(T\right)$ (see \eqref{27}).  To determine the optimal values of the system parameters that would implement the maximum possible number of free photons with decreasing the temperature, we shall involve numerical calculations. To do it, we introduce the dimensionless quantities

\begin{equation}\begin{split}\begin{gathered}
\label{30}
 \tau \equiv \frac{T}{\varepsilon _{\alpha _{2} } -\varepsilon _{\alpha _{1} } -\hbar \omega _{0} } ,
\ \ \ \ \
a \equiv \frac{n }{n_{ph}}-1,
\\
\theta \equiv \frac{1}{n_{ph} } \frac{g_{ph} g_{\alpha _{1} } }{g_{\alpha _{2} } } \left(\frac{m^{*}}{2\pi \hbar ^{2} } \right)^{3/2} \left|\varepsilon _{\alpha _{2} } -\varepsilon _{\alpha _{1} } -\hbar \omega _{0} \right|^{3/2} ,
\end{gathered}\end{split}\end{equation}

\noindent in terms of which the expression \eqref{28} for the density of free photons can be reduced to the form:

\begin{equation}\begin{split}\begin{gathered}
\label{31}
 \frac{\widetilde{n}_{ph} \left(\tau \right)}{n_{ph} } =-\frac{1}{2} \left\{a+\theta \left|\tau \right|^{3/2} \exp\left[\frac{1}{\tau} \right]\right\}+
\\
\frac{1}{2} \left\{a^{2} +2\left(a-2\right)\theta \left|\tau \right|^{3/2} \exp\left[\frac{1}{\tau} \right]+\theta ^{2}  \left|\tau \right|^{3} \exp\left[\frac{2}{\tau} \right]\right\}^{1/2}
\end{gathered}\end{split}\end{equation}

\begin{figure}[b]
\includegraphics[width=0.9\linewidth]{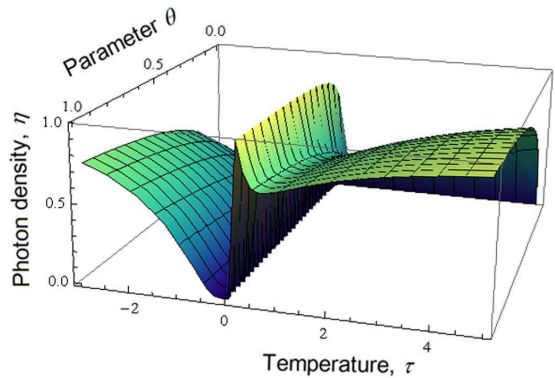}
\caption{
(Color online)
Dependence of the density of free photons in system (dimensionless quantity $\eta \equiv \widetilde{n}_{ph} / n_{ph} $, see \eqref{31}) on temperature $\tau$  and describing parameter  $\theta$, see \eqref{30}) plotted for $a=0.5$ (see also \eqref{30}).  The density of free photons increases with decreasing the temperature in the range $0<\tau<1/2$. This result does not depend on  $\theta$ and depends negligibly on $a$. Therefore, it leads to condition \eqref{33} (see also {32}), facilitating Bose condensation of photons in dilute nondegenerated gases of atoms.
}
\label{optimization}
\end{figure}

Figure 1 shows 3D plot of the ratio $\widetilde{n}_{ph} \left(\tau,\theta \right)/n_{ph} $ as a function of dimensionless parameters $\tau $ and $\theta $. This plot shows that the density of free photons increases with decreasing the temperature (dimensionless parameter $\left|\tau \right|$) in the range of $0<\tau <\frac{1}{2} $. Moreover, this tendency is observed almost  for all $\theta $.  Thus, analyzing the expression \eqref{31} (or \eqref{28}), we can conclude that if the physical characteristics of the system satisfy the relation

\begin{equation}
\label{32}
0<\frac{T}{\varepsilon _{\alpha _{2} } -\varepsilon _{\alpha _{1} } -\hbar \omega _{0} } <\frac{1}{2},
\end{equation}

\noindent then decreasing the temperature the number of free photons in system increases, that promotes the appearance of photons BEC. Here we must make the following remark. The presence of zero in the left-hand side of \eqref{32} is justified only in a formal sense. As temperature decreases, the system approaches to the critical point, where equations \eqref{22}, obtained under the assumptions that all the components of the system to be non-degenerated, $T\gg T_{c} $ (see (20)), become incorrect. For this reason, a temperature below the critical value in condition \eqref{22} cannot be considered. Due to this circumstance, the inequality \eqref{32} has to be written correctly in the following form:

\begin{equation}\begin{split}\begin{gathered}
\label{33}
\frac{\varepsilon _{\alpha _{2} } -\varepsilon _{\alpha _{1} } }{\hbar \omega _{0} } >1+\frac{2T}{\hbar \omega _{0} },
\ \ \ \
 T\gg T_{c}.
\end{gathered}\end{split}\end{equation}

Using solutions \eqref{26} and taking into account \eqref{27} - \eqref{28}, one can justify that in the case when \eqref{33} is satisfied, conditions \eqref{20} for all the three components to be nondegenerated are satisfied too.
It is necessary to note here the following circumstance. Expression (8) represents the closeness of the photon frequency   $\omega \left(\textbf{\textit{k}} \right) $ to the frequency  $\omega_{0}$ . If thus one assumes that the frequency  $\omega_{0}$   is close to the resonant frequency

\[\hbar \omega _{0} \approx \varepsilon _{\alpha _{2} } -\varepsilon _{\alpha _{1} } ,\]

\noindent then the implementation of condition \eqref{33} is only possible in the case of $\frac{T}{\hbar \omega _{0} } \ll 1$. Note also that this relation holds in the real conditions of current experiments on the condensation of photons \cite{11}-\cite{12}, where $\frac{T}{\hbar \omega _{0} } \simeq \frac{1}{80} $.

\section{Bose condensate of photons in dilute nondegenerated gas of atoms}
\label{photon condensate}

We now proceed to the study of equilibrium state of system with the presence of Bose condensate of photons. The most interesting case reveals when the BEC of photons appears in conditions of nondegeneration of atomic components in both  quantum states. In such a situation, conditions \eqref{20} should be written as:

\begin{equation}\begin{split}\begin{gathered}
\label{34}
\exp \left(\frac{\varepsilon _{\alpha _{1} } -\mu _{1} }{T} \right)\gg 1,
\ \ \ \ \
\exp \left(\frac{\varepsilon _{\alpha _{2} } -\mu _{2} }{T} \right)\gg 1,
\\
\exp \left(\frac{\hbar \omega _{0} -\mu ^{*} }{T} \right)\sim 1,
\end{gathered}\end{split}\end{equation}

\noindent therefore the first two of the expressions \eqref{21} remain valid. Then using \eqref{17}, equations \eqref{15} and \eqref{16}  can be reduced:

\begin{equation}\begin{split}\begin{gathered}
\label{35}
n=\left(\frac{mT}{2\pi \hbar ^{2} } \right)^{3/2} \exp \left(\frac{\mu _{2} }{T} \right)
\\
\times \left[g_{\alpha _{1} } \exp \left(-\frac{\varepsilon _{\alpha _{1} } +\mu ^{*} }{T} \right)
+g_{\alpha _{2} } \exp \left(-\frac{\varepsilon _{\alpha _{2} } }{T} \right)\right],
\\
n_{ph} =\frac{g^{*} }{2\pi ^{2} \hbar ^{3} } \int _{0}^{\infty }dp\frac{p^{2} }{\exp \left[\frac{\hbar \omega _{0} -\mu ^{*} +\left({p^{2} \mathord{\left/ {\vphantom {p^{2}  2m^{*} }} \right. \kern-\nulldelimiterspace} 2m^{*} } \right)}{T} \right]-1}
\\
+g_{\alpha _{2} } \left(\frac{mT}{2\pi \hbar ^{2} } \right)^{3/2} \exp \left(\frac{\mu _{2} }{T} \right)\exp \left(-\frac{\varepsilon _{\alpha _{2} } }{T} \right).
\end{gathered}\end{split}\end{equation}

\noindent Expressing the quantity $\exp \left({\mu _{2} \mathord{\left/ {\vphantom {\mu _{2}  T}} \right. \kern-\nulldelimiterspace} T} \right)$ from the first equation of \eqref{35} and substituting it into the second one, we obtain the following equation

\begin{equation}\begin{split}\begin{gathered}
\label{36}
\frac{g^{*} }{2\pi ^{2} \hbar ^{3} } \int _{0}^{\infty }dp\frac{p^{2} }{\exp \left[\frac{\hbar \omega _{0} -\mu ^{*} +\left({p^{2} \mathord{\left/ {\vphantom {p^{2}  2m^{*} }} \right. \kern-\nulldelimiterspace} 2m^{*} } \right)}{T} \right]-1}
\\
=n_{ph} -\frac{ng_{\alpha _{2} } \exp \left[-\frac{\varepsilon _{\alpha _{2} } }{T} \right]}{g_{\alpha _{1} } \exp \left[-\frac{\varepsilon _{\alpha _{1} } +\mu ^{*} }{T} \right]+g_{\alpha _{2} } \exp \left[-\frac{\varepsilon _{\alpha _{2} } }{T} \right]},
\end{gathered}\end{split}\end{equation}

\noindent which need to be supplemented by the equation \eqref{17} connecting the chemical potentials of all the three components. Recall that the left-hand side of the equation \eqref{36} is an expression for the density of free photons in the system. In this sense, the equation \eqref{36} is similar to the source equation for the study of BEC in an ideal gas of atoms, see for example \cite{14}. In other words, we have reduced the problem of Bose-Einstein condensation of photon gas, which is in thermodynamic equilibrium with nondegenerate atomic gas, to the problem of Bose-Einstein condensation in an ideal gas of photons with a certain effective density. Thus the right-hand side of equation \eqref{36} should be considered as such effective density.

 As is well known (see e.g. \cite{14}), for an ideal gas of Bose atoms with the binding energy $\varepsilon _{\alpha } $ in the temperature range $T\le T_{c} $, the chemical potential $\mu \left(T\right)$ must satisfy the condition $\mu \left(T\right)=\varepsilon _{\alpha } $. The same condition should define the transition temperature of BEC in ideal gas, $\mu \left(T_{c} \right)=\varepsilon _{\alpha } $. Thus, in the case of ideal gas of photons according to \eqref{36}, we have (see also \cite{13}):

\begin{equation}\begin{split}\begin{gathered}
\label{37}
\mu ^{*} \left(T\le T_{c} \right)=\hbar \omega _{0}.
\end{gathered}\end{split}\end{equation}

\noindent Taking into account \eqref{37} and following the procedure of  \cite{14} (see also \cite{13}), one obtains from \eqref{36} an intricate transcendental equation for the critical temperature $T_{c} $ as a function of the total density of photons $n_{ph} $ and atoms $n$ in the system,

\begin{equation}\begin{split}\begin{gathered}
\label{38}
\frac{g^{*} \left(2m^{*} T_{c} \right)^{3/2} }{4\pi ^{2} \hbar ^{3} } \Gamma \left(3/2\right)\zeta \left(3/2\right)
\\
=n_{ph} -n\left\{1+\frac{g_{\alpha _{1} } }{g_{\alpha _{2} } } \exp \left[\frac{\varepsilon _{\alpha _{2} } -\varepsilon _{\alpha _{1} } -\hbar \omega _{0} }{T_{c} } \right]\right\}^{-1} ,
\end{gathered}\end{split}\end{equation}

\noindent and the expression for the density of condensed photons $n_{ph}^{cond} \left(T\right)$  as a function of temperature $T<T_{c} $, for the same values of $n_{ph} $ and $n$:

\begin{equation}\begin{split}\begin{gathered}
\label{39}
n_{ph}^{cond} \left(T\right)=n_{ph} -n \left\{ 1+\frac{g_{\alpha _{1} } }{g_{\alpha _{2} } } \exp \left[\frac{\varepsilon _{\alpha _{2} } -\varepsilon _{\alpha _{1} } -\hbar \omega _{0} }{T} \right]\right\} ^{-1}
\\
-\frac{g^{*}}{4\pi ^{2} } \Gamma \left(3/2\right)\zeta \left(3/2\right) \left( \frac{2m^{*} T }{\hbar^{2}} \right)^{3/2} .
\end{gathered}\end{split}\end{equation}

Expression \eqref{37} yields both the chemical potential $\mu _{2} $ from the equation \eqref{35}

\begin{equation}\begin{split}\begin{gathered}
\label{40}
\exp \left[\frac{\mu _{2} }{T} \right]=n\left(\frac{mT}{2\pi \hbar ^{2} } \right)^{-3/2}
\\
\times  \left \{ g_{\alpha _{1} } \exp \left[-\frac{\varepsilon _{\alpha _{1} } +\hbar \omega _{0} }{T} \right]+g_{\alpha _{2} } \exp \left[-\frac{\varepsilon _{\alpha _{2} } }{T} \right]\right \} ^{-1},
\end{gathered}\end{split}\end{equation}

\noindent and, taking into account the equation \eqref{17} rewritten in form

\begin{equation}
\label{41}
\mu _{1} +\hbar \omega _{0} =\mu _{2},
\end{equation}

\noindent also the chemical potential $\mu _{1} $

\begin{equation}\begin{split}\begin{gathered}
\label{42}
\exp \left[\frac{\mu _{1} }{T} \right]=n\left(\frac{mT}{2\pi \hbar ^{2} } \right)^{-3/2} \exp \left[-\frac{\hbar \omega _{0} }{T} \right]
\\
\times \left \{ g_{\alpha _{1} } \exp \left[-\frac{\varepsilon _{\alpha _{1} } +\hbar \omega _{0} }{T} \right]+g_{\alpha _{2} } \exp \left[-\frac{\varepsilon _{\alpha _{2} } }{T} \right]\right \} ^{-1} .
\end{gathered}\end{split}\end{equation}

\noindent Thus, in accordance with \eqref{22}, the densities of atomic components are determined.

In some particular cases, the expression \eqref{38} for the critical temperature can be significantly simplified. In fact, if

\begin{equation}\begin{split}\begin{gathered}
\label{43}
0<\frac{\varepsilon _{\alpha _{2} } -\varepsilon _{\alpha _{1} } -\hbar \omega _{0} }{T_{c} } <1,
\\
\exp \left[\frac{\varepsilon _{\alpha _{2} } -\varepsilon _{\alpha _{1} } -\hbar \omega _{0} }{T_{c} } \right]\sim 1,
\end{gathered}\end{split}\end{equation}

\noindent the critical temperature, as is easily seen, is given by formula:

\begin{equation}\begin{split}\begin{gathered}
\label{44}
T_{c} \approx \frac{1}{2m^{*} } \left\{\frac{4\pi ^{2} \hbar ^{3} n_{ph}^{eff} }{g^{*} \Gamma \left(3/2\right)\zeta \left(3/2\right)} \right\}^{2/3} ,
\end{gathered}\end{split}\end{equation}

\begin{equation}\begin{split}\begin{gathered}
n_{ph}^{eff} \equiv \frac{n_{ph} \left(g_{\alpha _{1} } +g_{\alpha _{2} } \right)-ng_{\alpha _{2} } }{g_{\alpha _{1} } +g_{\alpha _{2} } } .
\end{gathered}\end{split}\end{equation}

\noindent As is easily seen, the expression for the critical temperature is identical by its form to the expression for the temperature of Bose-Einstein condensation in an ideal gas of ordinary Bose atoms with the mass $m^{*} $ and the density equal to $n_{ph}^{eff} $. In this limiting case, the transition temperature \eqref{44} depends on the density of atomic and photonic components, as well as on degeneracy multiplicity of the atomic levels. The obtained expression \eqref{44} for the temperature $T_{c} $ must satisfy the conditions \eqref{43}

\begin{equation}\begin{split}\begin{gathered}
\label{45}
0<2m^{*} \left(\varepsilon _{\alpha _{2} } -\varepsilon _{\alpha _{1} } -\hbar \omega _{0} \right)
\left\{\frac{g^{*} \; \Gamma \left(3/2\right)\zeta \left(3/2\right)}{4\pi ^{2} \hbar ^{3} n_{ph}^{eff} } \right\}^{2/3} <1. \end{gathered}\end{split}\end{equation}

\noindent This inequality links all the physical characteristics of the system under study.

In the case when parameters of structural units of the system satisfy relations

\begin{equation}\begin{split}\begin{gathered}
\label{46}
\frac{\varepsilon _{\alpha _{2} } -\varepsilon _{\alpha _{1} } -\hbar \omega _{0} }{T_{c} }<-1,
\\
 0<\exp \left[\frac{\varepsilon _{\alpha _{2} } -\varepsilon _{\alpha _{1} } -\hbar \omega _{0} }{T_{c} } \right]\ll 1, \end{gathered}\end{split}\end{equation}

\noindent the critical temperature is also given by the formula \eqref{44}, but with other expression for $n_{ph}^{eff} $

\begin{equation}\begin{split}\begin{gathered}
\label{47}
n_{ph}^{eff} =n_{ph} -n.
\end{gathered}\end{split}\end{equation}

\noindent It is easy to see that in this limiting case, the dependence on atomic levels degeneracy multiplicity vanishes. Furthermore, considering the necessity for condition $n_{ph} \sim n$ to be satisfied (see (19)), it is easy to conclude that the transition temperature in this case in accordance with \eqref{38}, \eqref{47} might be extremely low. However, it could be guessed also by examining Fig.1 for negative values of $\tau $.

 In the case of correctness of the inequalities

\begin{equation}\begin{split}\begin{gathered}
\label{48}
\frac{\varepsilon _{\alpha _{2} } -\varepsilon _{\alpha _{1} } -\hbar \omega _{0} }{T_{c} } >1,
\\
 \exp \left[\frac{\varepsilon _{\alpha _{2} } -\varepsilon _{\alpha _{1} } -\hbar \omega _{0} }{T_{c} } \right]\gg 1,
\end{gathered}\end{split}\end{equation}

\noindent the critical temperature is also given by theformula \eqref{44}, and one can easily convince that in this case effective density equals

\begin{equation}\begin{split}\begin{gathered}
\label{49}
n_{ph}^{eff} =n_{ph}.
\end{gathered}\end{split}\end{equation}

\noindent In this limiting case the dependences on the atomic component density, as well as on atomic degeneracy multiplicity, vanish. Again, in the last two cases it is also necessary for conditions \eqref{46} and \eqref{48} to be satisfied. In this case we get the relationship between the system physical parameters, similar to \eqref{45}.

It is necessary to resurrect here the question of the optimal relation between the parameters of the system structural units. We have shown previously (see \eqref{32} and \eqref{33}), that if $T\gg T_{c} $, the most advantageous is the condition $\left(\varepsilon _{\alpha _{2} } -\varepsilon _{\alpha _{1} } -\hbar \omega _{0} \right)>2T$, when with lowing the temperature the number of free photons in the system increases, that promotes the appearance of photons BEC at higher temperatures. The last case allows us to verify it directly. In fact, in this case the transition temperature in accordance with \eqref{44} and \eqref{49} is given by expression:

\begin{equation}\begin{split}\begin{gathered}
\label{50}
T_{c} \simeq \frac{\hbar ^{2}}{2m^{*} } \left\{\frac{4\pi ^{2}  n_{ph} }{g^{*} \; \Gamma \left(3/2\right)\zeta \left(3/2\right)} \right\}^{2/3}.
\end{gathered}\end{split}\end{equation}

This expression has the same appearance as if photon Bose condensation occurs in an ideal gas of photons in the total absence of atoms in the system. However, it is necessary to keep in mind that the relation $n_{ph} \sim n$ (see (19)) has to be satisfied always , since we are in the two-level atoms framework. Nevertheless, assuming that in all considered above limiting cases (see  \eqref{43} - \eqref{50}), the density $n_{ph} $ is the same, then in the last case \eqref{48}, the critical temperature, given by expression \eqref{50}, will be the highest.

 Starting from \eqref{40}-\eqref{42} one can also verify that the relations \eqref{34}, which determine the conditions of atomic components to be nondegenerated, are always satisfied. The reason, ensuring their implementation, is the existence of a strong inequality $m^{*} \ll m$.

Expressions \eqref{38} and \eqref{39} allow solving the inverse problem, i.e. what should be the relationship between the physical parameters of the system for photon condensation to occur for given, , and desirably "high", temperatures?  The term "high temperature" in the present article assumes the temperature when atomic components are degenerated, including, for example, the range of room temperatures. For this purpose, we consider again the expression \eqref{38}, rearranging it to the form:

\begin{equation}\begin{split}\begin{gathered}
\label{51}
n_{ph} =\frac{g^{*} \left(2m^{*} T_{c} \right)^{3/2} }{4\pi ^{2} \hbar ^{3} } \Gamma \left(3/2\right)\zeta \left(3/2\right)
\\
+n  \left\{1+\frac{g_{\alpha _{1} } }{g_{\alpha _{2} } } \exp \left[\frac{\varepsilon _{\alpha _{2} } -\varepsilon _{\alpha _{1} } -\hbar \omega _{0} }{T_{c} } \right]\right\}^{-1}.
\end{gathered}\end{split}\end{equation}

As an example, we examine now the solutions of announced problem in two (may be said extremely different) cases. In the first case, we consider conditions of appearance of Bose condensate of photons that are in the thermodynamic equilibrium with gases, which have parameters that are close to those of ultracold vapors of alkali metals. In the second case, we consider the same issue for photons that are in equilibrium with "classic" ideal gas of atoms at room temperature.

To begin, we examine the possibility of appearance of photon BEC in the first case. To do it, one has to assign the values of density and temperature in \eqref{51}. Choosing these parameters, it is natural to focus on their values in real experiments of trapped BEC in alkali metal vapors \cite{1}-\cite{3}. For this reason, we shall assume $n\sim 10^{13} cm^{-3} $ and $T\sim 10^{-5} K$. For such temperatures and densities, the vapors of alkali metals should remain nondegenerate. One can estimate the first term on the right-hand side of \eqref{51}, assuming $m^{*} \sim 10^{-33} g$ (as in \cite{11})

\[\frac{g^{*} \left(2m^{*} T_{c} \right)^{3/2} }{4\pi ^{2} \hbar ^{3} } \Gamma \left(3/2\right)\zeta \left(3/2\right)\sim 1\; cm^{-3} .\]

\noindent Estimating the second term, we can use one of the basic requirements of the present article, namely, the interrelation $n_{ph} \sim n$ to be correct. Then from \eqref{51} one obtains:

\[n_{ph} \sim 1\; cm^{-3}
\\
 +\frac{10^{13} cm^{-3}}{1+\frac{g_{\alpha _{1} } }{g_{\alpha _{2} } } \exp \left[ \left( \varepsilon _{\alpha _{2} } -\varepsilon _{\alpha _{1} } -\hbar \omega _{0} \right) / T_{c}  \right]}  .\]

\noindent Thus, to satisfy the condition $n_{ph} \sim n$, it is necessary to satisfy the following relation:

\begin{equation}\begin{split}\begin{gathered}
\label{52}
\frac{\left|\varepsilon _{\alpha _{2} } -\varepsilon _{\alpha _{1} } -\hbar \omega _{0} \right|}{T_{c} } \sim 1.
\end{gathered}\end{split}\end{equation}

\noindent Recall that the condition $n_{ph} \sim n$ was introduced in order to justify the use of the assumption of two-level atoms (see (19)). As it follows from these estimates, the contribution of the first term in \eqref{51} can be neglected. Thus, one concludes that the Bose condensation of photons in ultracold vapors of alkali metals can be possible if the following approximate equation is held:

\begin{equation}\begin{split}\begin{gathered}
\label{53}
n_{ph} \approx n\left\{1+\frac{g_{\alpha _{1} } }{g_{\alpha _{2} } } \exp \left[\frac{\varepsilon _{\alpha _{2} } -\varepsilon _{\alpha _{1} } -\hbar \omega _{0} }{T_{c} } \right]\right\}^{-1} .
\end{gathered}\end{split}\end{equation}

 We now consider the possibility of Bose condensation of photon gas that is in thermodynamic equilibrium with an ideal gas of atoms at room temperature, $T\sim 300K$. We assume that the density of the atomic subsystem is close to the density of the "classical" ideal gas, that is $n\sim {\rm 1}0^{{\rm 19}} cm^{-3} $. Again, for these conditions one can estimate the first term in \eqref{51} as

\[\frac{g^{*} \left(2m^{*} T_{c} \right)^{3/2} }{4\pi ^{2} \hbar ^{3} } \Gamma \left(3/2\right)\zeta \left(3/2\right)\sim 10^{11} \, cm^{-3} .\]

\noindent As in the previous case, estimating the second term we shall keep in mind the condition $n_{ph} \sim n$. Then from \eqref{51} one gets:

\[n_{ph} \sim 10^{11} \, cm^{-3} +\frac {10^{19} cm^{-3} } {1+\frac{g_{\alpha _{1} } }{g_{\alpha _{2} } } \exp \left[ \left( \varepsilon _{\alpha _{2} } -\varepsilon _{\alpha _{1} } -\hbar \omega _{0} \right) / T_{c}  \right]}  .\]

\noindent Thus, as in the previous case, it turns out that the requirement $n_{ph} \sim n$ leads to the relation \eqref{52}. Furthermore, as in the previous case, the contribution of the first term in \eqref{51} can be neglected, and again one arrives to the expression \eqref{53}.

 The result is really amazing, taking into account the fact how significantly the systems under consideration in two cases differ. Of course, it should be understood that the result of the above calculations depends on the value of the photon effective mass $m^{*} $, which in various systems may differ significantly from the value $6.7\cdot 10^{-33} g$, given in \cite{11}. However, we should emphasize here that the value $m^{*} \sim 10^{-33} g$, taken for above numerical estimations, corresponds to the most interesting case of the visible light, i.e. wavelength range $\lambda \in \left\{380 \ nm;750 \  nm\right\}$. As was mentioned above, the effective mass of photon is determined by the spectrum cut-off frequency $\omega _{0} $ and the speed of propagation of photons $v$ in the medium (see \eqref{9}). These values also determine the form of photon's dispersion law in the medium (see in this connection \eqref{7} and \eqref{8}). Therefore, mentioned above estimation results clearly indicate the fact that there are conditions under which one can neglect the contribution of the "traditional" for ideal gas term in the expression for the critical temperature (i.e. the first term on the right-hand side of\eqref{51}). Consequently, there should be physical systems, similar to those ones considered before, where expression for the transition temperature differs cardinally from the "traditional" dependence for the case of ideal gas of ordinary atoms. In the next section, we shall consider such situation in details.
In conclusion of this section we emphasize that condition \eqref{52} is not exotic. For example, for  $\hbar \omega_{0} \approx 2.1 \ eV$      \cite{11} and room temperature $T \sim 300 \ K$, condition (53) is satisfied for Sodium line corresponding to the transition between $3P_{3/2}$ and $3S_{1/2}$ (see \cite{alkalidata}).

\section{Mechanism of fast condensation of photons in dilute nondegenerate atomic gas}
\label{fast condensation}

To examine the issue that had risen in the previous section, let us return to the equation \eqref{51} being rearranged as follows:

\begin{equation}\begin{split}\begin{gathered}
\label{54}
\frac{n_{ph} }{n} =\Gamma \left(3/2\right)\zeta \left(3/2\right)\frac{g^{*} \left(2m^{*} T_{c} \right)^{3/2} }{4\pi ^{2} \hbar ^{3} n}
\\
+\left \{1+\frac{g_{\alpha _{1} } }{g_{\alpha _{2} } } \exp \left[\frac{\varepsilon _{\alpha _{2} } -\varepsilon _{\alpha _{1} } -\hbar \omega _{0} }{T_{c} } \right]\right \}^{-1}.
\end{gathered}\end{split}\end{equation}

As was noted earlier, the first term in \eqref{54} corresponds to the traditional character of Bose condensate behavior that occurs gaining the phase transition temperature, defined, for example, by the expression \eqref{50} when the condition \eqref{48} is true. This character of Bose condensation has been observed experimentally as in ultracold gases of alkali atoms (see e.g. \cite{1}-\cite{5}), as well in systems with conserved number of photons \cite{11}-\cite{12}. The second term of \eqref{54} gives a significantly different behavior of the critical temperature of Bose condensation of photons in dilute nondegenerate gas of atoms. We shall reveal this fact in details.

	We stress once again that \eqref{54} is a transcendental equation with regard to the temperature of Bose condensation $T_{c} $, and it can not be solved in elementary functions. On this account, in Section 4 we were only interested in the various limiting cases, which (in our opinion) are close to the current experimental conditions for the photon condensation in the presence of matter \cite{11}. However, since the system under study differs significantly from the system of work \cite{11}, one should also consider the limiting case, when the second term of \eqref{54} is dominant. Some estimates that indicate the legitimacy of the existence of such a regime were given in the end of the previous section. In this limiting case, the expression \eqref{54} simplifies

\begin{equation}\begin{split}\begin{gathered}
\label{55}
\frac{n_{ph} }{n} \simeq \left \{1+\frac{g_{\alpha _{1} } }{g_{\alpha _{2} } } \exp \left[\frac{\varepsilon _{\alpha _{2} } -\varepsilon _{\alpha _{1} } -\hbar \omega _{0} }{T_{c} } \right]\right \}^{-1},
\\n_{ph} <n.
\end{gathered}\end{split}\end{equation}

We specify here that in this section we consider positive values of the energies difference $\varepsilon _{\alpha _{2} } -\varepsilon _{\alpha _{1} } -\hbar \omega _{0} $, see in this regard comments to the formula \eqref{47} and Fig.1. Accordingly, the critical temperature of the photon Bose condensate phase formation in the system under study for this regime is given by

\begin{equation}\begin{split}\begin{gathered}
\label{56}
T_{c} =\left(\varepsilon _{\alpha _{2} } -\varepsilon _{\alpha _{1} } -\hbar \omega _{0} \right) Ln^{-1} \left[\frac{g_{\alpha _{2} } }{g_{\alpha 1} } \frac{n-n_{ph} }{n_{ph} } \right].
\end{gathered}\end{split}\end{equation}

Note that the expression \eqref{56} reveals the logarithmic dependence of Bose condensation temperature in the system under study on the densities of its components. This behavior significantly differs from traditional power dependence (for example, see \eqref{50}). In this case, the feature of the logarithmic regime of condensation is the following circumstance. With a proper pumping the system with photons, the critical temperature of Bose condensation in accordance with formula \eqref{56} can unlimitedly increase within a finite density of photons. From a physical point of view, it means that in this regime the Bose condensation of photons can take place much faster than in mentioned earlier case of the power dependence \eqref{50}. In this regard, it makes sense to talk about the mechanism of fast condensation of photons within the conditions of the regime \eqref{56}. As far as we know, previously in literature this kind of dependence was not mentioned.

One can easily see that providing the sharp increase of photon Bose condensation critical temperature, the argument of logarithm in \eqref{56} should tend to unity:

\begin{equation}\begin{split}\begin{gathered}
\label{57}
\frac{g_{\alpha _{2} } }{g_{\alpha _{1} } } \frac{\left(n-n_{ph} \right)}{n_{ph} }  \to 1,
\ \ \ \ \ \ \
T_{c} \to \infty,    \end{gathered}\end{split}\end{equation}

\noindent that implies the need to fulfill the following relation (see \eqref{55}):

\begin{equation}\begin{split}\begin{gathered}
\label{58}
n_{ph} \to n\cdot \frac{g_{\alpha _{2} } }{g_{\alpha _{1} } +g_{\alpha _{2} } } .
\end{gathered}\end{split}\end{equation}

\noindent Note that the expression \eqref{58} satisfies the requirement $n_{ph} \sim n$. In other words, the number of photons, pumped into the system, is sufficiently large to provide macroscopically large number of free photons in system. Thus, in accordance with \eqref{58} the amount of pumped photons should be less than the total number of atoms, therefore in real experiment the probability of excitation of higher atomic levels is extremely small. This circumstance can justify the application of the two-level atoms model.

We study now in details the transition from the traditional for Bose condensation "power"  regime \eqref{50} to the fast regime of photons condensation \eqref{56}. To do it, one has to examine the expression \eqref{54} in the region, where no small parameters exist. For this reason we use the numerical methods as auxiliary. To this end, we introduce the following dimensionless quantities:

\begin{equation}\begin{split}\begin{gathered}
\label{59}
\tau \equiv \frac{T_{c} }{\varepsilon _{\alpha _{2} } -\varepsilon _{\alpha _{1} } -\hbar \omega _{0} } ,
\ \ \ \ \
  \eta \equiv \frac{n_{ph} }{n},
\ \ \ \ \
g_{12}  \equiv \frac{g_{\alpha _{1} } }{g_{\alpha _{2} } },
\\
\gamma \equiv \frac{g^{*} \, \Gamma \left(3/2\right)\, \, \zeta \left(3/2\right)}{\sqrt{2} \pi ^{2} \hbar ^{3} n} \left[m^{*} \left(\varepsilon _{\alpha _{2} } -\varepsilon _{\alpha _{1} } -\hbar \omega _{0} \right)\right]^{3/2}.
\end{gathered}\end{split}\end{equation}

\noindent In terms of these dimensionless parameters the equation \eqref{54} can be reduced to the form:

\begin{equation}\begin{split}\begin{gathered}
\label{60}
\gamma  \left|\tau \right|^{3/2} +\left(1+g_{12} \exp \left(1/\tau \right)\right)^{-1} =\eta.
\end{gathered}\end{split}\end{equation}

\noindent Recall again that in accordance with the results, obtained in Section~\ref{symmetric phase}, the region of negative $\tau $ is not very promising from the viewpoint of BEC implementation at high temperatures. Since we have agreed in this section to assume $\left(\varepsilon _{\alpha _{2} } -\varepsilon _{\alpha _{1} } -\hbar \omega _{0} \right)>0$, the absolute value sign in \eqref{60} can be omitted.

 Figure 2 shows the results of numerical solution of the equation \eqref{60}, reflecting the dependences $\tau \left(\eta \right)$ for different values of the system parameter $\gamma $. These plots, in accordance with the definitions \eqref{59}, reveal the dependence of the critical temperature of photons Bose condensation on the density of pumped into the system photons. To simplify calculations, we assumed $g_{\alpha _{1}} / g_{\alpha _{2} }  =1$. The power law dependence (lower solid line) corresponds to the numerical solution of equation \eqref{60} with $\gamma =1$, i.e. in the case when the first term of equation \eqref{54} dominates. Logarithmic regime of condensation (upper solid line) occurs for values $\gamma \sim 10^{-2} $ and lower, i.e. in the case when the first term in equation \eqref{60} (also see \eqref{54}) can be neglected. Intermediate regimes between "power" and "logarithmic", for lucidity, are shown by dashed curves.

The results of numerical solution witness the transition to the fast regime of Bose condensation of photons if the parameters of system satisfy the relation $\gamma \lesssim 10^{-2}$ (see \eqref{59}). Therefore, the following condition is in fact imposed on the quantities that characterize the system under study

\begin{figure}[b]
\includegraphics[width=0.95\linewidth]{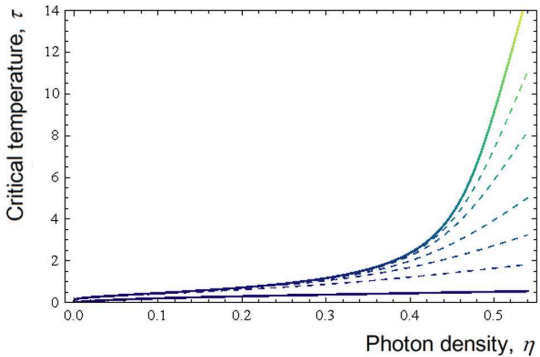}
\caption{
(Color online)
Dependence of Bose condensation critical temperature $\tau$ on the density of pumped into the system photons  $\eta$ for different regimes (i.e. for different characteristics of system $\gamma$), see notations of variables \eqref{59}. For simplicity, all the curves are shown for  $g_{\alpha _{1}} / g_{\alpha _{2} }  =1$. The lowest solid line corresponds to the true "power" regime \eqref{50}, the highest solid line is for the true "logarithmic" regime \eqref{56}. Intermediate regimes for different values of $\gamma$ are  shown in dashed lines.
}
\label{optimization}
\end{figure}

\begin{equation}\begin{split}\begin{gathered}
\label{61}
\frac{1}{n\hbar ^{3} } \left[m^{*} \left(\varepsilon _{\alpha _{2} } -\varepsilon _{\alpha _{1} } -\hbar \omega _{0} \right)\right]^{3/2} \ll 1.
\end{gathered}\end{split}\end{equation}

\noindent In addition, as can be seen from \eqref{55}, the requirement $n_{ph} \sim n$ implies the condition $\left(\varepsilon _{\alpha _{2} } -\varepsilon _{\alpha _{1} } -\hbar \omega _{0} \right)\sim T_{c} $. Thus, the condition for realization the  fast condensation regime \eqref{61} can be  also written as

\begin{equation}\begin{split}\begin{gathered}
\label{62}
\kappa \equiv \frac{1}{n} \left(\frac{T_{c} m^{*} }{\hbar ^{2} } \right)^{3/2} \ll 1.
\end{gathered}\end{split}\end{equation}

\noindent Note that this condition can be achieved in dilute gases at rather high temperatures, for example, room temperatures. Thus, for example, for the same photon mass $m^{*} \sim 10^{-33} g$ (see \cite{11}), room temperature ($T=300K$), and the density of diluted gas $n=10^{16} cm^{-3} $, one obtains $\kappa \sim 10^{-5} $. As it is seen, the condition \eqref{62} is satisfied very well, even up to characteristic temperatures of some types of plasma.

 One can easily show that the conditions \eqref{34} are also satisfied. In fact, there are no doubts in validity of conditions for atomic components to be nondegenerated, because they are valid even in the case of lower temperatures, determined by power condensation regime \eqref{50}. As for the conditions for photonic component to be degenerate, it is implied by the correctness of equality $\mu ^{*} =\hbar \omega _{0} $ (see \eqref{37}).

It is clear from the above analysis that the achievement of the "high temperature" regime of Bose condensation of photons is essentially defined by both the total density of atomic components and the density of the total amount of photons, pumped into the system. Thus, one can affect the achievement of the transition point both with the change of atoms number in system, and with pumping photons with a help of laser. Naturally, the second option seems much more practical from the point of view of experiments. In this regard, the following question rises: what is the efficiency of additional pumping of photons in the system under study? As a quantitative characteristic of such efficiency it is natural to choose the quantity $dT_{c} / dn_{ph} $. It should be emphasized that we do not actually proceed to the dynamic characteristics of the system. It is assumed that after some controlled pumping of additional photons in the system (for example, laser impulse pumping), the system requires some time to come to the thermodynamic equilibrium state. In this sense, we will call the introduced quantity $dT_{c} / dn_{ph} $ the efficiency of the given regime of Bose condensation of photons.

In the case of logarithmic dependence of critical temperature on the total photon density (see \eqref{56}), one can readily come to the following expression:

\begin{equation}\begin{split}\begin{gathered}
\label{63}
\frac{dT_{c} }{dn_{ph} } = \frac{n \left(\varepsilon _{\alpha _{2} } -\varepsilon _{\alpha _{1} } -\hbar \omega _{0} \right)}{n_{ph} \left(n-n_{ph} \right)}
 Ln^{-2} \left[\frac{g_{\alpha _{2} } }{g_{\alpha 1} } \frac{n-n_{ph} }{n_{ph} } \right].
\end{gathered}\end{split}\end{equation}

In the case, when the "power law" condensation mechanism (given by formula (50)) predominates, one obtains

\begin{equation}\begin{split}\begin{gathered}
\label{64}
\frac{dT_{c} }{dn_{ph} } =\frac{1}{3m^{*} } \left\{\frac{4\pi ^{2} \hbar ^{3} }{g^{*} \, \Gamma \left(3/2\right)\zeta \left(3/2\right)} \right\}^{2/3} n_{ph}^{-1/3} \propto n_{ph}^{-1/3}.
\end{gathered}\end{split}\end{equation}

Comparing expressions \eqref{63} and \eqref{64} it becomes obviously that the efficiency of power condensation regime decreases monotonically with increasing the density of pumped photons, whereas the efficiency of logarithmic condensation regime can increase under certain conditions. One should simultaneously keep in mind that the requirement $n_{ph} \sim n$ must be held.

 For a more detailed consideration of the introduced quantity we will again involve numerical methods. To do it, we use the general expression \eqref{54} as the equation, defining an implicit dependence of the critical temperature of Bose condensation $T_{c} $ on the total density $n_{ph} $ of pumped into the system photons. Taking a derivative from the both sides of \eqref{54} with respect to $n_{ph} $, after some hackneyed transformations one can obtain:

\begin{equation}\begin{split}\begin{gathered}
\label{65}
\frac{d\tau }{d\eta } =\left\{\frac{3}{2} \gamma  \tau ^{1/2} + \frac{g_{12} \exp \left[1 / \tau  \right]}{\tau ^{2} \left(1+g_{12} \exp \left[1 / \tau  \right]\right)^{2}}  \right\}^{-1}.
\end{gathered}\end{split}\end{equation}

Here one should use one more time the definitions \eqref{59}. The obtained expression \eqref{65} should be considered in couple with \eqref{60} as a system of equations. This system allows to determine both the critical temperature of Bose condensation $\tau $, and the efficiency of chosen regime (in the sense that was defined above) being characterized by quantity ${dT_{c} \mathord{\left/ {\vphantom {dT_{c}  dn_{ph} }} \right. \kern-\nulldelimiterspace} dn_{ph} } $, as the functions of the pumped photons density $\eta $ for given parameters $\gamma $ and $g_{12} $.

Figure 3 shows plots for the dependences $\tau '_{\eta } \left(\eta \right)$ as a result of the numerical solution of equations \eqref{65} and \eqref{60} for different values of $\gamma $. As in Figure 2, the power and logarithmic regimes are shown by the solid lines (bottom and top ones, respectively), intermediate regimes are shown by dashed lines. The plot shows that the efficiency of logarithmic condensation regime \eqref{63} increases with the number of photons while the efficiency of power condensation regime \eqref{64} decreases. Therefore, the curves in Figure 3 give an understanding how with an appropriate selection of system parameters (see \eqref{59}) one can switch the system under study to the desired regime with any given condensation rate. Note also that at low densities of pumped photons, both dependences \eqref{63} - \eqref{64} behave themselve in a similar way. Thus, the numerical results witness in favour of the predominant implementation of the fast logarithmic regime of photons Bose condensation in a wide range of pumped into the system photons densities.

\begin{figure}[t]
\includegraphics[width=0.9\linewidth]{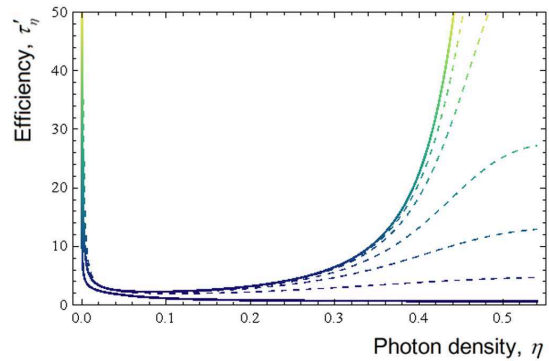}
\caption{
(Color online)
Dependence of the photon pumping efficiency $\tau '_{\eta } \left(\eta \right)$  (see \eqref{65}) on the density of pumped into the system photons  $\eta$ for different system characteristics  $\gamma$, see notations \eqref{59}. For simplicity, all curves are given for $g_{\alpha _{1}} / g_{\alpha _{2} }  =1$. The lowest solid line corresponds to the true "power" regime \eqref{50}, the highest solid line is for the true "logarithmic" regime \eqref{56}. Intermediate regimes for different values of $\gamma$ are  shown in dashed lines. Minimum point for logarithmic regime \eqref{63} corresponds to  $\eta \simeq 0.08$.
}
\label{optimization}
\end{figure}

In other words, trying to achieve experimentally a state with a BEC of photons, that are  in thermodynamic equilibrium with an ideal gas of effectively two-level atoms with the help of set of methods, similar to that described in \cite{11}-\cite{12}, the logarithmic condensation regime occurs with the overwhelming probability. Moreover, as it was already noted, it applies to a wide range of physical systems, that are in equilibrium with photons: from ultracold gases of alkali metals to some certain types of an ideal plasma (for plasma classification types and their featured parameters see e.g. \cite{20}).

\section{Conclusion}

The main purpose of the present work, as it has been already noted, was to reveal the possibility of BEC existence in a gas of photons that are in the state of thermodynamic equilibrium with the gas of atoms that can repeatedly absorb and reemit photons. Such a mechanism of gaining the thermodynamic equilibrium was supposed in the present paper to be dominating. The main reason for this consideration is the desire to have some resemblance between the system under study and the system of \cite{11}-\cite{12}, where photon BEC was first obtained experimentally. We managed to demonstrate the possibility of achieving such a state of the photon component within a framework of fairly simple model, indicating, however, the conditions under which the situation could be close to real one. In some cases, our model requires improvements or corrections. It is connected, for example, with taking into account that Bose condensation of photons is affected by interactions between the atoms of atomic components and photon scattering by atoms. This is a separate, sophisticated issue that in our view is beyond the scope of the present paper. Its solution has to be in framework of the microscopic approach based on the low-temperature quantum electrodynamics. In our view, a very promising is the approach proposed in \cite{21}. Authors are currently working on these issues.

The atoms of atomic components in the present paper are considered to be two-level. It is a fairly widespread technique in theoretical physics, optics, and photonics. But it should be noted that from the standpoint of this paper, it is not a fundamental limitation. It only simplifies  calculations, allowing in some cases to obtain the results in analytical form. The equations  \eqref{15} -\eqref{17} can be formulated for an arbitrarily large number of components. But in this case there rises a challenge of a controlled numerical solution of such equations.

	It is also worth to discuss the problem of photon dispersion in matter. The importance of solving this problem is related, first of all, to the fact that the explicit form of  $\omega \left(\textbf{\textit{k}}\right)$ determines the effective mass of a photon in the matter (see (7) - (13)). In this paper  we have used for numerical estimations the value of the effective mass of the photon from \cite{11}. However, the effective mass of photons in different substances can significantly vary. It can be related to the interaction features of photons with the structural units of a particular substance. For this reason, ideally the dependence of the photon frequency on the wave vector $\omega \left(\textbf{\textit{k}}\right)$  has to be calculated independently for each specific system in a self-consistent way. Moreover, such self-consistent calculations of the photon dispersion in matter must proceed from the first principles underlying the description of quantum many-particle systems. In other words, in the case of issues related to the description of BEC of photons, calculations of photon dispersion in matter have to be based on nonrelativistic quantum electrodynamics and quantum statistical mechanics. In its most general form, this problem seems to be unreasonable. However, in some cases, it is quite realistic. It was successfully accomplished, for example, in \cite{17}-\cite{18} for the description of slowing and absorption processes of electromagnetic waves in gases of alkali metals with BEC or propagation of  relativistic charged particles in the same substances \cite{19}. These studies used a microscopic approach based on the low energy quantum electrodynamics, constructed in \cite{21}. There is a hope to generalize the approach of \cite{21}, \cite{17}-\cite{19} in order to calculate the effective mass of the photon in the systems similar to those considered in the present paper.

In conclusion, the authors would also emphasize that mentioning an ideal plasma in our work is related to the demonstration of the vastness of the temperature range in which  used approach and obtained results are valid, but it is not a statement about the direct applicability of these results to photons in plasma. For a description of such processes in an ideal plasma, our model requires a substantial modification (in this regard see \cite{9}).

%

%
%
\end{document}